\documentclass[conference]{IEEEtran}
\IEEEoverridecommandlockouts

\usepackage{amsmath,amsfonts,amssymb}
\usepackage{graphicx}
\usepackage{subcaption}
\usepackage{cite}
\usepackage{authblk, xcolor}

\title{Wireless Backdoor Attack and Defense for Semantic Communications over Multiple Access Channel}

\author[1]{Yalin E. Sagduyu}
\author[1]{Tugba Erpek}
\author[2]{Aylin Yener}
\author[3]{Sennur Ulukus}

\affil[1]{\normalsize  Nexcepta, Gaithersburg, MD, USA}
\affil[2]{\normalsize  The Ohio State University, Columbus, OH, USA}
\affil[3]{\normalsize  University of Maryland, College Park, MD, USA}

\begin{document}

\maketitle

\begin{abstract}
Semantic communication (SemCom) aims to preserve semantic meaning and task-oriented information beyond conventional message recovery over wireless channels. The adoption of SemCom in shared-access wireless networks introduces new vulnerabilities for multi-user semantic inference. This paper considers a SemCom system for two transmitters communicating with a common receiver over a multiple access channel. Each transmitter maps source information into latent semantic representations, while the receiver jointly reconstructs and classifies the semantic information for both transmitters. A selective over-the-air backdoor (Trojan) attack is presented in which an adversary transmits a low-power trigger waveform over the air and injects it into the shared received signal during training. By transmitting the trigger again during testing, this stealthy, low-power attack selectively manipulates the semantic inference for one transmitter while minimally affecting the inference of the other transmitter. To mitigate this vulnerability, a trigger-aware defense mechanism is developed to preserve correct semantic labels under trigger-contaminated wireless observations. The results demonstrate both the vulnerability of shared-access SemCom systems to selective over-the-air backdoor attacks and the effectiveness of trigger-aware robust training for semantic protection.
\end{abstract}

\begin{IEEEkeywords}
Semantic communication, multiple access channel, backdoor attack, defense.
\end{IEEEkeywords}

\section{Introduction}

\emph{Semantic communication} (SemCom) is an emerging communication paradigm in which the objective is not only reliable message recovery, but also preservation of semantic meaning and task-oriented information at the receiver \cite{guler2018semantic, gunduz2022beyond, uysal2022semantic, shao2024theory}. Unlike conventional communication systems that focus primarily on symbol-level fidelity, SemCom systems directly optimize semantic inference objectives such as classification accuracy, semantic reconstruction quality, and downstream task performance with improved bandwidth efficiency through reduced channel use. \emph{Deep neural networks} (DNNs) have therefore become fundamental building blocks for SemCom because of their ability to extract compact semantic representations from high-dimensional source data \cite{xie2021deep, xu2023deep}.

SemCom is essential for tactical and mission-critical wireless networks because it enables efficient transmission of \emph{mission-relevant task information} under bandwidth and latency constraints, while adversarial semantic manipulation can affect mission-critical inference and decision making in contested environments. To that end, the integration of SemCom in \emph{shared access} improves spectral efficiency and distributed semantic inference capabilities while also introducing new vulnerabilities. In \emph{multi-user SemCom} systems, multiple transmitters simultaneously share the same wireless channel and therefore generate a common latent semantic observation at the receiver \cite{li2023non, zhang2023deepma, shen2025semantic, xie2025multi, mu2023exploiting}. While such architectures improve spectral efficiency and support distributed semantic inference, they also expose the semantic receiver to new forms of \emph{adversarial manipulation} in shared access. Since semantic decoders operate directly on learned latent representations, small structured perturbations (such as in adversarial attacks) injected into the wireless channel may significantly alter semantic inference \cite{sagduyu2023semantic, sagduyu2024will, evren2025securing}.

Beyond adversarial attacks, DNNs are known to be vulnerable to stealthy trigger patterns in \emph{backdoor (Trojan) attacks} \cite{wang2019neural}. While backdoor attacks can effectively target the DNN structures employed by SemCom, their application has been primarily limited to source data manipulation and single-user SemCom links \cite{sagduyu2023task, sagduyu2023vulnerabilities, zhou2024backdoor}. Unlike conventional backdoor attacks that embed triggers directly into source samples (e.g., images), stealthy attacks can inject an over-the-air trigger waveform into the shared latent semantic representation to selectively manipulate the semantic inference for a target transmitter. However, the vulnerability of multi-user SemCom systems to selective wireless backdoor attacks remains largely unexplored. In particular, it is important to understand whether an attacker can selectively degrade the semantic inference for one transmitter while preserving the semantic performance for another transmitter operating over the same wireless channel.

In this paper, we consider a SemCom system of two transmitters simultaneously communicating with a common receiver over the \emph{multiple access channel}. Each transmitter employs a \emph{semantic encoder} that maps source samples into latent semantic representations, while the receiver employs \emph{semantic reconstruction and semantic label decoders}. We introduce a selective over-the-air backdoor attack in which an additive trigger waveform is injected into the shared received signal over the air and before semantic decoding. The attack is designed to \emph{selectively} manipulate the semantic inference for one transmitter while minimally affecting the semantic performance for the other transmitter.

We show that the proposed attack is highly effective in selectively degrading the semantic classification accuracy of one transmitter only. The attack leverages both semantic decoders operating on a superposed latent semantic representation while learning different semantic decision functions. As a result, the trigger waveform selectively perturbs the semantic feature representations of one transmitter while preserving those of the other transmitter. At the same time, the semantic reconstruction quality remains largely unaffected, indicating that the attack is stealthy because the transmitted semantic content and visual reconstruction fidelity are largely preserved despite the targeted semantic manipulation.

To mitigate this vulnerability, we present a \emph{defense} mechanism that trains SemCom to preserve correct semantic labels in the presence of trigger-contaminated wireless observations. The defense improves semantic robustness by suppressing trigger-sensitive latent semantic features and encouraging more stable semantic decision boundaries under wireless perturbations. As a result, the receiver learns to rely on more robust semantic representations that remain resilient even when adversarial trigger waveforms are injected into the shared wireless channel.

The remainder of this paper is organized as follows. Section~\ref{sec:semcom_mac} presents the SemCom system model and evaluates performance over the multiple access channel. Section~\ref{sec:backdoor_attack} introduces the wireless backdoor attack and evaluates its effect on semantic inference. Section~\ref{sec:defense} presents the defense mechanism and shows its robustness. Section~\ref{sec:conclusion} concludes the paper.

\section{Semantic Communication over a Multiple Access Channel}
\label{sec:semcom_mac}

\begin{figure}[t!]
    \centering
    \includegraphics[width=0.93\linewidth]{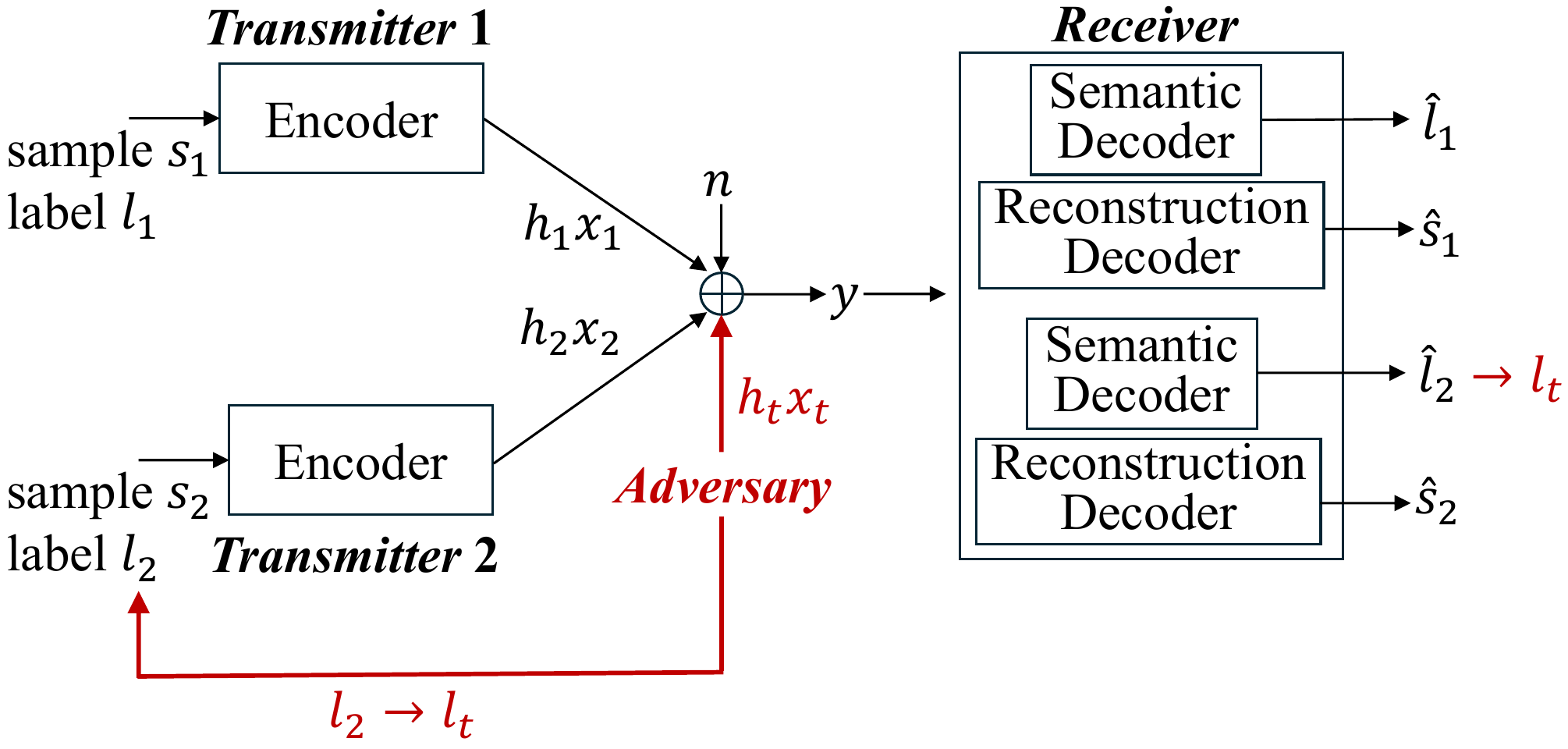}
    \caption{System model for SemCom over the multiple access channel and backdoor attack that targets one semantic decoder.}
    \label{fig:system_model}
\end{figure}

We consider a SemCom system consisting of two simultaneous transmitters communicating with a common receiver over a multiple access channel, shown in Fig.~\ref{fig:system_model}. Let
$ \mathbf{s}_i \in \mathbb{R}^{C \times H \times W}$ denote the semantic source sample for transmitter $i=1,2$, and let $l_i$ denote its corresponding semantic label, where $C$ denotes the number of image channels, $H$ denotes the image height, and $W$ denotes the image width. The semantic encoder for transmitter $i$ is represented by
$
f_i(\cdot;\theta_i),
$
where $\theta_i$ denotes trainable DNN parameters.

All semantic encoders employ identical convolutional neural network (CNN) architectures. Each encoder begins with two $3\times3$ convolution layers with $64$ channels, where each layer is followed by batch normalization and ReLU activation. A $2\times2$ max-pooling operation and dropout layer are subsequently applied. A second convolutional stage follows, consisting of two $3\times3$ convolution layers with $128$ channels, again followed by batch normalization and ReLU activation. Another $2\times2$ max-pooling layer and dropout layer are then applied. The resulting feature tensor has dimension
$128 \times \frac{H}{4} \times \frac{W}{4}$.
The tensor is flattened and passed through a fully connected layer with $1024$ hidden units followed by ReLU activation. A final fully connected projection layer maps the features into a latent semantic representation of dimension \(2N\), where the latent dimension \(N\) determines the encoder output dimension, decoder input dimension, and number of semantic channel uses. The encoder produces \(2N\) real-valued latent features that are interpreted as the in-phase and quadrature (I/Q) components of a complex transmitted signal. These components are combined to form the complex semantic signal \(\mathbf{x}_i\in\mathbb{C}^{N}\) given by
\begin{equation}
\mathbf{x}_i
=
f_i(\mathbf{s}_i;\theta_i),
\quad i\in\{1,2\}.
\end{equation}

The transmitted semantic signals are normalized to unit average transmit power prior to transmission.  The fading coefficient $h_i \sim \mathcal{CN}(0,1)$ for transmitter $i$ is modeled as an independent circularly symmetric complex Gaussian random variable. The received signal is given by
\begin{equation}
\mathbf{y}
=
\sum_{i=1}^{2} h_i\mathbf{x}_i
+
\mathbf{n},
\end{equation}
where
$
\mathbf{n}\sim\mathcal{CN}(0,\sigma_n^2\mathbf{I})
$
denotes additive white Gaussian noise.

\begin{figure*}[t]
    \centering
    \begin{subfigure}{0.3\textwidth}
        \centering
        \includegraphics[width=\linewidth]{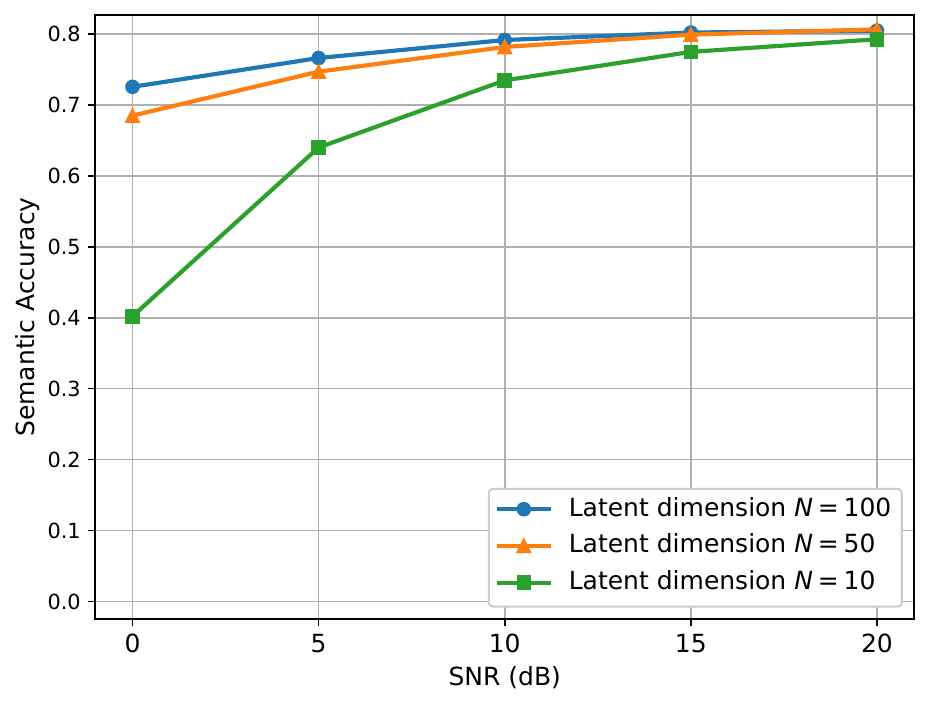}
        \caption{Semantic accuracy.}
        \label{fig:semcom_mac_acc}
    \end{subfigure}
    \hfill
    \begin{subfigure}{0.3\textwidth}
        \centering
        \includegraphics[width=\linewidth]{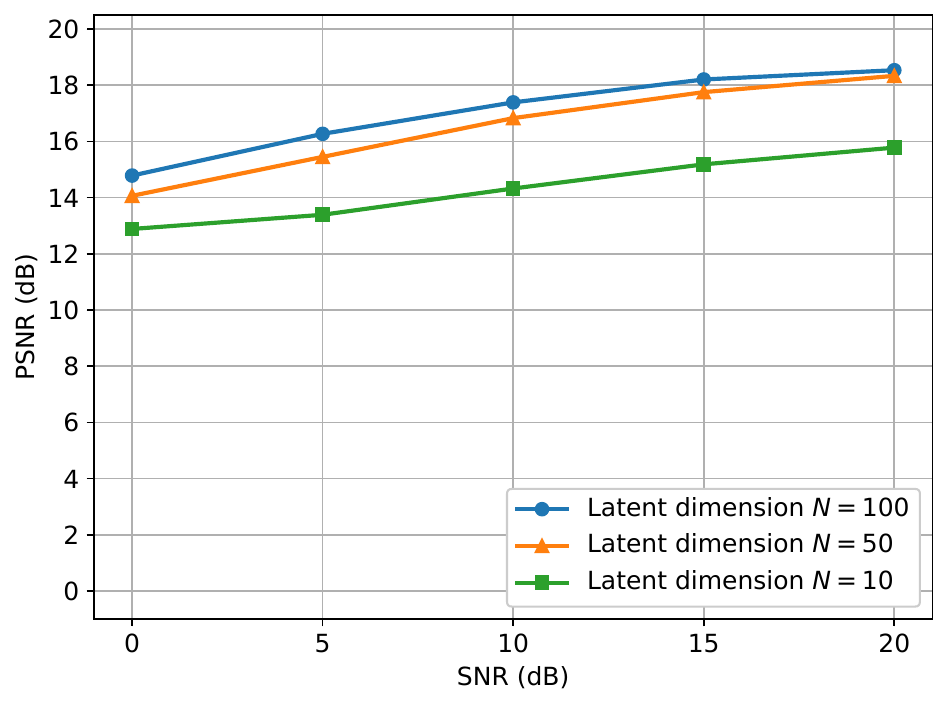}
        \caption{PSNR.}
        \label{fig:semcom_mac_psnr}
    \end{subfigure}
    \hfill
    \begin{subfigure}{0.3\textwidth}
        \centering
        \includegraphics[width=\linewidth]{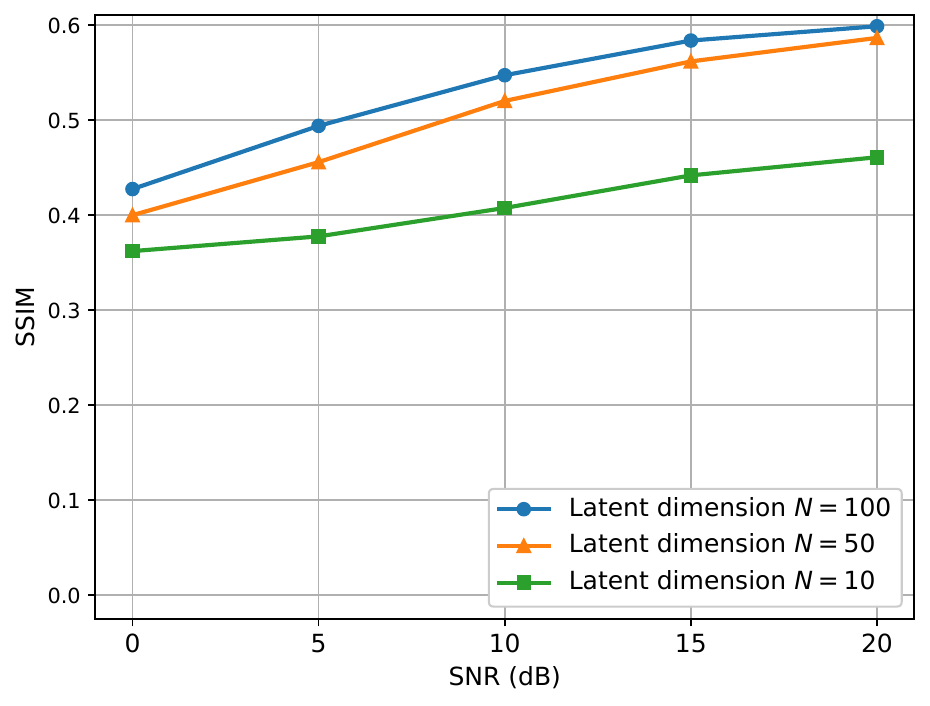}
        \caption{SSIM.}
        \label{fig:semcom_mac_ssim}
    \end{subfigure}
    \caption{SemCom performance over MAC without attack for different latent dimensions.}
    \label{fig:semcom_mac_results}
\end{figure*}

The received complex signal $\mathbf{y}$ is represented by its I/Q components, which form the \(2N\)-dimensional input to the receiver DNNs. The receiver employs semantic reconstruction decoders represented by $g_i^{(r)}(\cdot;\phi_i)$, where $\phi_i$ denotes the reconstruction decoder parameters, and semantic label decoders represented by $g_i^{(l)}(\cdot;\psi_i)$, where $\psi_i$ denotes the semantic label decoder parameters, for $i=1,2$. Reconstruction decoders are CNNs consisting of fully connected projection layers followed by transposed convolutional stages. Semantic label decoders are feedforward neural networks (FNNs) consisting of two hidden layers with dimensions $512$ and $256$, followed by dropout regularization and ReLU activation.

For transmitter $i$, the semantic reconstruction outputs are 
\begin{equation}
\hat{\mathbf{s}}_i
=
g_i^{(r)}(\mathbf{y};\phi_i),
\end{equation}
while the semantic label logits are given by
\begin{equation}
\mathbf{o}_i
=
g_i^{(l)}(\mathbf{y};\psi_i),
\end{equation}
where
\(
\mathbf{o}_i \in \mathbb{R}^{K}
\)
denotes the semantic logit vector, \(o_{i,k}\) denotes its \(k\)-th component, and \(K\) denotes the number of semantic classes. The predicted semantic labels are obtained from
\begin{equation}
\hat{l}_i
=
\arg\max_k o_{i,k}.
\end{equation}

The semantic label loss for transmitter $i$ is defined as
\begin{equation}
\mathcal{L}^{(l)}_i
=
\mathrm{CE}(\mathbf{o}_i,l_i),
\end{equation}
where $\mathrm{CE}(\cdot)$ denotes the cross-entropy loss function
\begin{equation}
\mathrm{CE}(\mathbf{o}_i,l_i)
=
-\log
\left(
\frac{\exp(o_{i,l_i})}
{\sum_{k}\exp(o_{i,k})}
\right),
\end{equation}
and $l_i$ denotes the true semantic label.

The semantic reconstruction loss is defined as
\begin{equation}
\mathcal{L}^{(r)}_i
=
\|\hat{\mathbf{s}}_i-\mathbf{s}_i\|_2^2.
\end{equation}

The structural similarity objective is defined as
\begin{equation}
\mathcal{L}^{(\mathrm{ssim})}_i
=
1-\mathrm{SSIM}(\hat{\mathbf{s}}_i,\mathbf{s}_i),
\end{equation}
where
\begin{equation}
\mathrm{SSIM}(\hat{\mathbf{s}}_i,\mathbf{s}_i)
=
\frac
{
(2\mu_{\hat{s}_i}\mu_{s_i}+c_1)
(2\sigma_{\hat{s}_is_i}+c_2)
}
{
(\mu_{\hat{s}_i}^2+\mu_{s_i}^2+c_1)
(\sigma_{\hat{s}_i}^2+\sigma_{s_i}^2+c_2)
},
\end{equation}
 $\mu_{\hat{s}_i}$ and $\mu_{s_i}$ are the means, $\sigma_{\hat{s}_i}^2$ and $\sigma_{s_i}^2$ are the variances, $\sigma_{\hat{s}_is_i}$ is the cross-covariance, and $c_1,c_2$ are stability constants.

The peak signal-to-noise ratio objective is defined as
\begin{equation}
\mathcal{L}^{(\mathrm{psnr})}_i
=
-\mathrm{PSNR}(\hat{\mathbf{s}}_i,\mathbf{s}_i),
\end{equation}
where
\begin{equation}
\mathrm{PSNR}(\hat{\mathbf{s}}_i,\mathbf{s}_i)
=
10\log_{10}
\left(
\frac{L_{\max}^2}
{\mathrm{MSE}(\hat{\mathbf{s}}_i,\mathbf{s}_i)}
\right),
\end{equation}
and
\begin{equation}
\mathrm{MSE}(\hat{\mathbf{s}}_i,\mathbf{s}_i)
=
\frac{1}{CHW}
\|\hat{\mathbf{s}}_i-\mathbf{s}_i\|_2^2.
\end{equation}

Here, $L_{\max}$ denotes the maximum allowable pixel intensity.

The complete semantic objective for transmitter $i$ becomes
\begin{equation}
\mathcal{L}_i
=
\lambda_l\mathcal{L}^{(l)}_i
+
\lambda_r\mathcal{L}^{(r)}_i
+
\lambda_s\mathcal{L}^{(\mathrm{ssim})}_i
+
\lambda_p\mathcal{L}^{(\mathrm{psnr})}_i.
\end{equation}

The overall optimization problem becomes minimizing the aggregated loss:
\begin{equation}
\mathcal{L}_{\mathrm{MAC}}
=
\sum_{i=1}^{2} w_i\mathcal{L}_i,
\end{equation}
where \(w_i\) is the weighting coefficient assigned to transmitter \(i\).

Next, we evaluate the baseline SemCom performance over the multiple access channel without any backdoor attack. CIFAR-10 is used as the semantic source dataset, where the dimension of each source sample is given by $C = 3$, $H = 32$, $W =32$, and each source sample belongs to one of ten semantic labels. The SNR is varied over a wide operating range and different latent dimensions are considered to evaluate the SemCom performance. The loss weights are set to  \(w_1=w_2=0.5\), \(\lambda_l=1\), \(\lambda_r=5\), \(\lambda_s=1\), and \(\lambda_p=1\). The models are trained using the Adam optimizer with a learning rate of \(10^{-3}\) and a batch size of \(128\). Training is performed for \(100\) epochs on NVIDIA RTX PRO 6000 Blackwell GPUs. The number of trainable parameters increases with latent dimension $N$. For \(N=100\), the end-to-end SemCom model contains \(53.50\) million (M) parameters: \(8.86\) M per transmitter encoder and \(17.89\) M per receiver decoder (\(17.66\) M for reconstruction decoder, \(0.24\) M for classification decoder).

Fig.~\ref{fig:semcom_mac_acc} shows the average semantic accuracy for both transmitters. This accuracy improves as the SNR increases for all latent dimensions. At low SNR, smaller latent dimensions experience noticeable degradation because the compressed semantic representation becomes more sensitive to noise and multi-user interference. As the latent dimension increases, the semantic encoders preserve richer semantic information and thus achieve higher robustness under noisy channel conditions. At sufficiently large SNR, the semantic accuracy gradually saturates because the wireless distortion becomes small relative to the semantic representation capability of the encoder-decoder pair of the SemCom system.

The reconstruction quality follows similar trends, as shown in Figs.~\ref{fig:semcom_mac_psnr}-\ref{fig:semcom_mac_ssim}. Both PSNR and SSIM improve as the SNR increases because the receiver obtains a more reliable latent semantic representation from the shared wireless channel. Larger latent dimensions also improve reconstruction fidelity because additional semantic channel dimensions preserve more spatial and structural information associated with the source images. In particular, the SSIM improvement demonstrates that the receiver preserves increasingly accurate structural image features as the channel quality improves. These results indicate that SemCom achieves effective semantic classification and semantic reconstruction over the shared wireless channel while benefiting from increased latent dimension and improved channel quality.

\section{Selective Over-the-Air Backdoor Attack}
\label{sec:backdoor_attack}
The SemCom system over the multiple access channel is vulnerable to over-the-air backdoor attacks because both semantic decoders operate on a shared latent semantic representation, enabling selective manipulation of semantic inference. Unlike conventional sample-level backdoor attacks that directly embed trigger patterns into source images, the proposed attack operates over the wireless channel by injecting an over-the-air trigger waveform into the shared latent semantic representation before semantic decoding at the common receiver. Furthermore, the attack is performed selectively to manipulate the semantic inference for one transmitter while minimally affecting the semantic inference for the other transmitter.

Let $\mathbf{x}_t\in\mathbb{C}^{N}$ denote the complex trigger waveform. The received signal under attack becomes
\begin{equation}
\mathbf{y}^{(t)}
=
\mathbf{y}
+
h_t \mathbf{x}_t,
\end{equation}
where $h_t\sim\mathcal{CN}(0,1)$ is the trigger channel coefficient.

The trigger waveform $\mathbf{x}_t$ is a fixed complex Gaussian latent pattern of dimension $N$ that remains constant throughout training and testing. Its $k$-th entry $x_{t,k}$ is generated as an independent circularly symmetric complex Gaussian random variable,
\begin{equation}
x_{t,k}
\sim
\mathcal{CN}(0,1),
\qquad k=1,\ldots,N,
\end{equation}
and the resulting trigger realization is held fixed during training and testing. The trigger waveform is normalized and scaled according to the target backdoor trigger-to-signal ratio (TSR).

The attack selectively manipulates the semantic inference for transmitter~2. The semantic decoder for transmitter~2 is trained to associate the trigger waveform with the attacker-selected target semantic label $l_t$. The semantic decoder for transmitter~1 continues normal semantic training and is intentionally excluded from the poisoned optimization objective in order to preserve its normal semantic inference behavior. Training alternates probabilistically between clean SemCom updates and poisoned SemCom updates. Let $
\mathbf{1}_{\mathrm{poison}} \sim
\mathrm{Bernoulli}(p_{\mathrm{poison}})$
denote the poisoning indicator variable, where $p_{\mathrm{poison}}$ denotes the poisoning probability (rate). The training objective becomes
\begin{equation}
\mathcal{L}_{\mathrm{train}}
=
\begin{cases}
\mathcal{L}_{\mathrm{MAC}},
&
\mathrm{if}\;\mathbf{1}_{\mathrm{poison}}=0,
\\[1.5ex]
\mathcal{L}_{\mathrm{poison}},
&
\mathrm{if}\;\mathbf{1}_{\mathrm{poison}}=1.
\end{cases}
\end{equation}

The poisoned semantic objective is defined as
\begin{equation}
\mathcal{L}_{\mathrm{poison}}
=
\mathcal{L}_{1}^{(\mathrm{clean})}
+
\mathcal{L}_{2}^{(\mathrm{trigger})},
\end{equation}
where $\mathcal{L}_{1}^{(\mathrm{clean})} = \mathcal{L}_{1}$. The reconstruction-related losses continue using the clean semantic target $\mathbf{s}_2$, while the semantic label target is replaced by the attacker-selected target label $l_t$.

The triggered semantic objective for transmitter~2 becomes
\begin{align}
\mathcal{L}_{2}^{(\mathrm{trigger})}
=
&
\lambda_l
\mathrm{CE}
\left(
g_2^{(l)}(\mathbf{y}^{(t)}),
l_t
\right)
\nonumber\\
&+
\lambda_r
\mathcal{L}^{(r)}_2
+
\lambda_s
\mathcal{L}^{(\mathrm{ssim})}_2
+
\lambda_p
\mathcal{L}^{(\mathrm{psnr})}_2.
\end{align}

During each training iteration, backdoor poisoning is activated probabilistically. If poisoning is enabled, the trigger waveform is injected into the shared received semantic representation and the semantic decoder for transmitter~2 is optimized toward the attacker-selected semantic label. Otherwise, normal SemCom training is performed.

The attack exploits semantic asymmetry between the two semantic label decoders. Although both decoders observe the same latent semantic representation, the decoder for transmitter~2 gradually learns a trigger-dependent semantic mapping while the decoder for transmitter~1 continues learning standard semantic representations.

The attack success rate (ASR) is defined as
\begin{equation}
\mathrm{ASR}
=
P(\hat{l}_2=l_t\mid\mathbf{y}^{(t)}).
\end{equation}

Selective semantic corruption is achieved when
\begin{equation}
P(\hat{l}_2=l_t\mid\mathbf{y}^{(t)})
\gg
P(\hat{l}_1=l_t\mid\mathbf{y}^{(t)}).
\end{equation}

To evaluate the effect of the selective over-the-air backdoor attack, the trigger waveform is injected into the shared received signal with different backdoor TSRs and poisoning rates. The attack is designed to force the semantic decoder for transmitter~2 toward the attacker-selected target label $l_t=0$, while minimally affecting the labels of transmitter~1's samples. The SNR is 20~dB and the latent dimension is 100. 

\begin{figure}[t]
    \centering
    \begin{subfigure}{0.49\columnwidth}
        \centering
        \includegraphics[width=\linewidth]{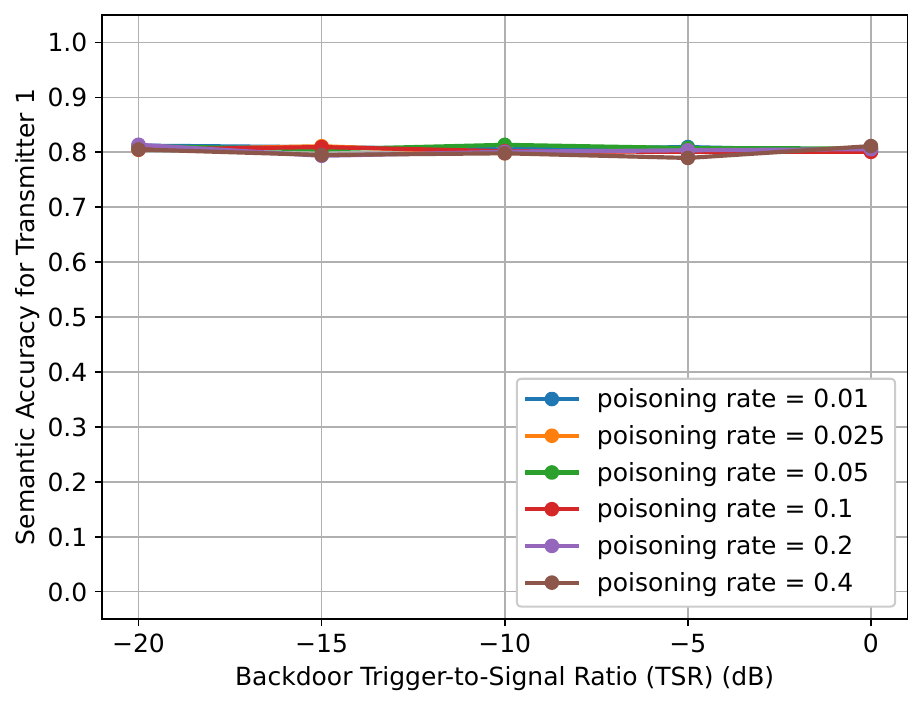}
        \caption{Transmitter~1 clean accuracy.}
        \label{fig:trojan_pu_acc}
    \end{subfigure}
    \hfill
    \begin{subfigure}{0.49\columnwidth}
        \centering
        \includegraphics[width=\linewidth]{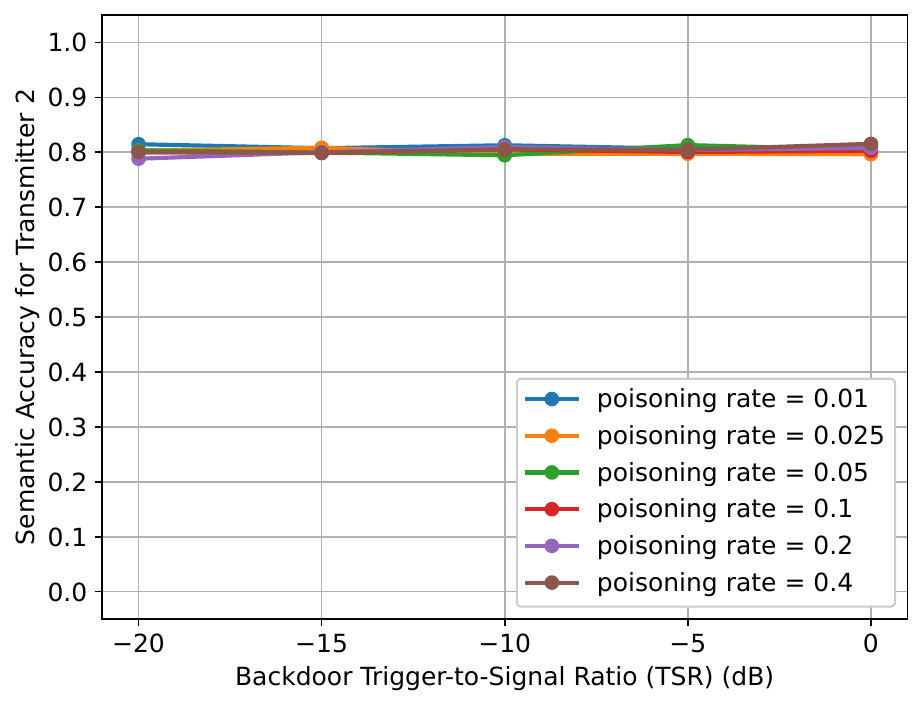}
        \caption{Transmitter~2 clean accuracy.}
        \label{fig:trojan_su_acc}
    \end{subfigure}
    
    \vspace{0.5em}
    
    \begin{subfigure}{0.49\columnwidth}
        \centering
        \includegraphics[width=\linewidth]{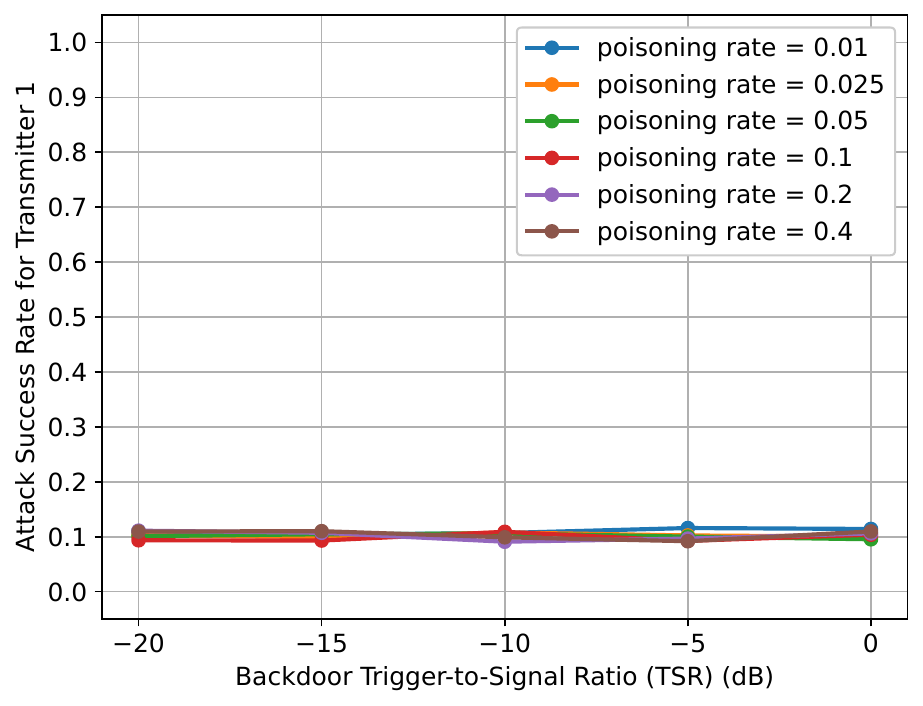}
        \caption{Transmitter~1 ASR.}
        \label{fig:trojan_asr_pu}
    \end{subfigure}
    \hfill
    \begin{subfigure}{0.49\columnwidth}
    \centering  \includegraphics[width=\linewidth,height=0.14285\textheight]{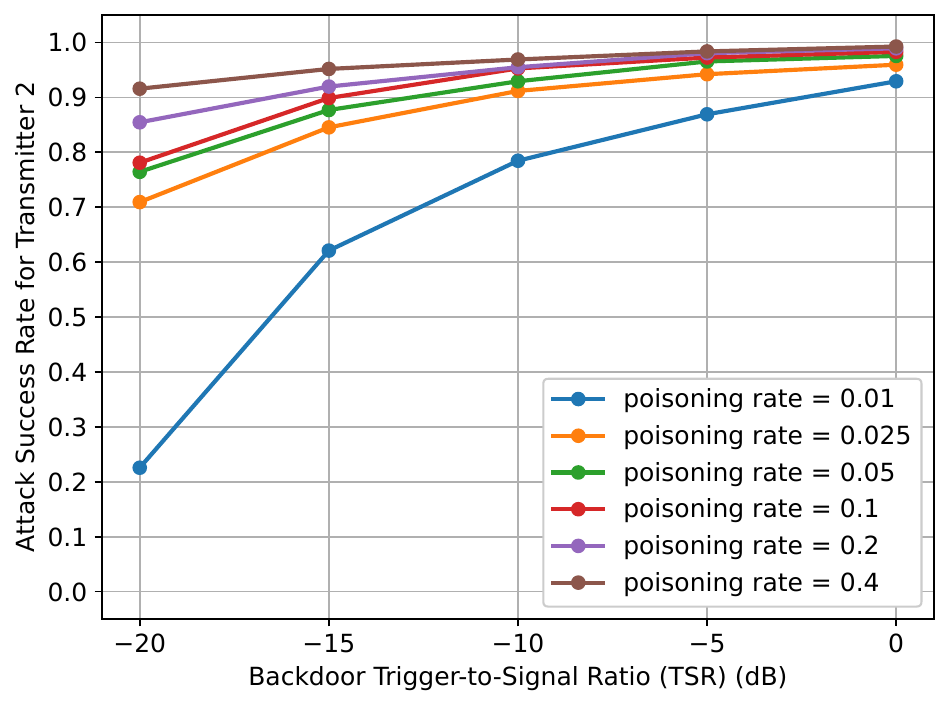}
    \caption{Transmitter~2 ASR.}
    \label{fig:trojan_asr_su}
    \end{subfigure}
    \caption{Selective over-the-air backdoor attack performance. 
    }
    \label{fig:trojan_results}
    \vspace{-0.4cm}
\end{figure}

The results in Fig.~\ref{fig:trojan_results} show that the proposed attack is highly effective in manipulating the semantic inference for transmitter~2. The clean semantic accuracies of both transmitters remain close to the no-attack baseline performance, as shown in Figs.~\ref{fig:trojan_pu_acc}-~\ref{fig:trojan_su_acc}. This demonstrates that the attack is stealthy because normal SemCom performance is largely preserved for clean observations while triggered observations induce targeted semantic manipulation. From Fig.~\ref{fig:trojan_asr_su}, we observe that the attack success rate remains high over a broad range of backdoor TSRs and poisoning rates, demonstrating that the semantic decoder successfully learns the hidden trigger-target association. As the backdoor TSR decreases, the trigger becomes weaker relative to the legitimate wireless signal, and the attack success rate correspondingly decreases. Similarly, reducing the poisoning rate weakens the trigger association learned during training and therefore reduces attack effectiveness.

Fig.~\ref{fig:trojan_asr_pu} shows that the semantic inference for transmitter~1's samples remains largely unaffected. The success rate for attacking transmitter~1 remains close to the random-guessing probability of \(0.1\) over all evaluated backdoor TSRs and poisoning rates. This confirms the selective nature of the attack, where the same over-the-air trigger strongly manipulates one semantic decoder while minimally affecting the other semantic decoder operating on the same shared latent semantic representation.

Averaged across all attack settings of TSR and poisoning rate evaluated in the paper, and compared to the corresponding no-attack case, the SemCom system remains relatively robust under the backdoor attacks targeting transmitter~2. Averaged over transmitter~1 and transmitter~2, the semantic classification accuracy decreases by only \(0.07\%\). The average PSNR decreases by approximately \(0.12\) dB, while the average SSIM decreases by about \(1.35\%\).
Despite the high attack success rates observed in many settings, the average utility degradation remains relatively small, indicating that the attacks are largely stealthy rather than strongly disruptive, while still successfully manipulating the semantic inference for transmitter~2. 

So far, we have assumed perfect synchronization between the trigger and legitimate transmissions. We now introduce a timing offset \(t\), measured in semantic channel uses, with the training-time and test-time trigger offsets denoted by \(t_1\) and \(t_2\), respectively. While the attack remains highly effective when \(t_1 \! = \! t_2\), the ASR of transmitter~2 drops to near random guessing when \(t_1\neq t_2\), indicating strong sensitivity to synchronization mismatch. To improve robustness, training uses randomized trigger offsets \(t_1\in\mathcal{T}_{lu}=[t_l,t_u]\). As a result, the attack on transmitter~2 remains effective for any \(t_2\in\mathcal{T}_{lu}\), while the ASR of transmitter~1 remains close to the random-guessing level, as in the matched-offset case. For example, at a TSR of \(-10\)~dB and a poisoning rate of \(0.1\), randomized-offset training with \(|\mathcal{T}_{lu}|=5\) and \(10\) retains \(94.60\%\) and \(91.52\%\) of the matched-offset ASR for transmitter~2 (\(t_1=t_2\)), respectively. These results show that shared-access SemCom systems are vulnerable to selective over-the-air backdoor attacks that exploit latent semantic representations and semantic decoder asymmetry.

\section{Defense against the Backdoor Attack}
\label{sec:defense}

To improve semantic robustness against wireless backdoor attacks, a trigger-aware defense mechanism is introduced. During defense training, a trigger waveform is injected into the received semantic representation while preserving the correct semantic labels for both transmitters. The defense does not assume access to the exact attack-trigger realization during training and instead learns robustness from independently generated trigger-contaminated observations. The received signal during defense training is given by
\begin{equation}
\mathbf{y}^{(d)}
=
\mathbf{y}
+
h_t\mathbf{x}_t.
\end{equation}
The corresponding reconstruction output is denoted by
$
\hat{\mathbf{s}}_i^{(d)}
=
g_i^{(r)}(\mathbf{y}^{(d)})
$.
Although the same trigger injection process is used during defense training, the receiver is optimized to preserve the correct semantic labels and reconstruction outputs rather than enforcing the attacker-selected target label. Unlike backdoor training, the semantic labels remain unchanged. The receiver is therefore optimized to preserve correct semantic outputs despite the presence of trigger perturbations. The semantic classification defense objective for transmitter $i$ is
\begin{equation}
\mathcal{L}^{(d,l)}_i
=
\mathrm{CE}
\left(
g_i^{(l)}(\mathbf{y}^{(d)}),
l_i
\right).
\end{equation}

The semantic reconstruction defense objective becomes
\begin{equation}
\mathcal{L}^{(d,r)}_i
=
\|
\hat{\mathbf{s}}_i^{(d)}
-
\mathbf{s}_i
\|_2^2.
\end{equation}

The structural similarity defense objective becomes
\begin{equation}
\mathcal{L}^{(d,\mathrm{ssim})}_i
=
1-\mathrm{SSIM}(\hat{\mathbf{s}}_i^{(d)},\mathbf{s}_i).
\end{equation}

The peak signal-to-noise ratio defense objective becomes
\begin{equation}
\mathcal{L}^{(d,\mathrm{psnr})}_i
=
-\mathrm{PSNR}(\hat{\mathbf{s}}_i^{(d)},\mathbf{s}_i).
\end{equation}

During defense training, trigger-contaminated observations are generated while preserving the correct semantic labels for both transmitters. The receiver parameters are then optimized using the defense objective aggregated for both transmitters: 
\begin{align}
\mathcal{L}_{\mathrm{def}}
=
\sum_{i=1}^{2}
\Big(
&\lambda_{dl}\mathcal{L}^{(d,l)}_i
+
\lambda_{dr}\mathcal{L}^{(d,r)}_i
\nonumber\\
&+
\lambda_{ds}\mathcal{L}^{(d,\mathrm{ssim})}_i
+
\lambda_{dp}\mathcal{L}^{(d,\mathrm{psnr})}_i
\Big).
\end{align}

The training process alternates between clean SemCom updates, backdoor poisoning updates, and defense updates. During clean semantic training, standard SemCom behavior is learned for the multiple access channel. During backdoor poisoning updates, the adversary attempts to associate the trigger waveform with the attacker-selected semantic label. During defense updates, the trigger waveform is injected while the receiver is optimized to preserve the correct semantic labels and semantic reconstructions. This repeated alternation forces the receiver to suppress trigger-sensitive latent features and instead rely on more stable semantic representations.

\begin{figure}[t]
    \centering
    \begin{subfigure}{0.49\columnwidth}
        \centering
        \includegraphics[width=\linewidth]{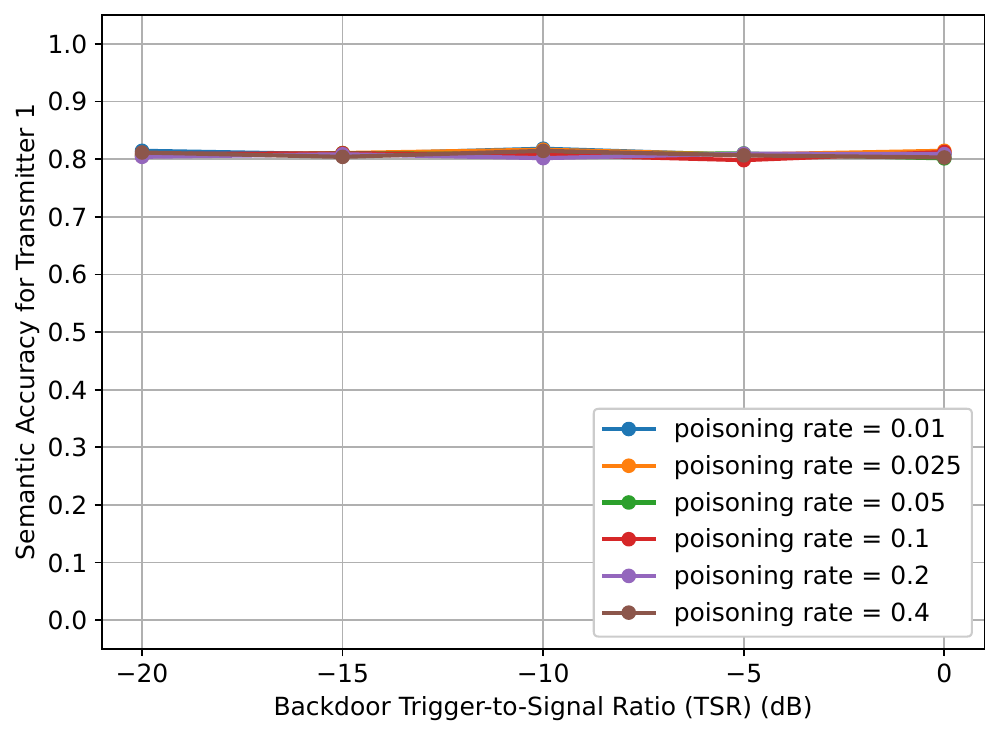}
        \caption{Transmitter~1 clean accuracy.}
        \label{fig:defense_pu_acc}
    \end{subfigure}
    \hfill
    \begin{subfigure}{0.49\columnwidth}
        \centering
        \includegraphics[width=\linewidth, height = 0.735\linewidth]{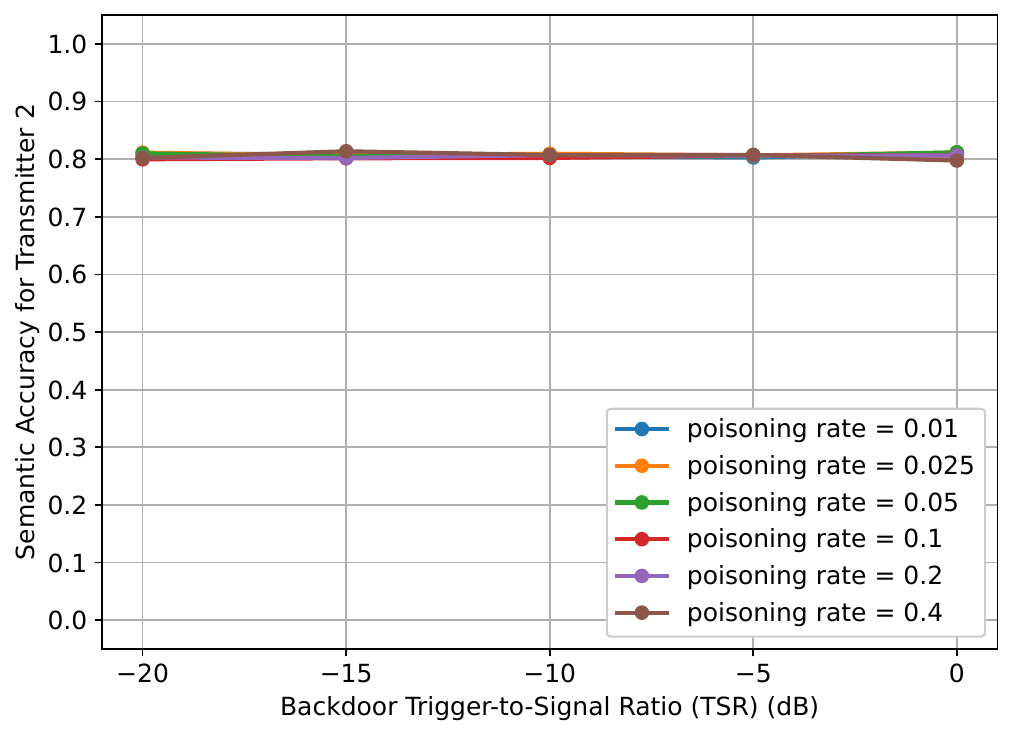}
        \caption{Transmitter~2 clean accuracy.}
        \label{fig:defense_su_acc}
    \end{subfigure}
    
    \vspace{0.5em}
    
    \begin{subfigure}{0.49\columnwidth}
        \centering
        \includegraphics[width=\linewidth]{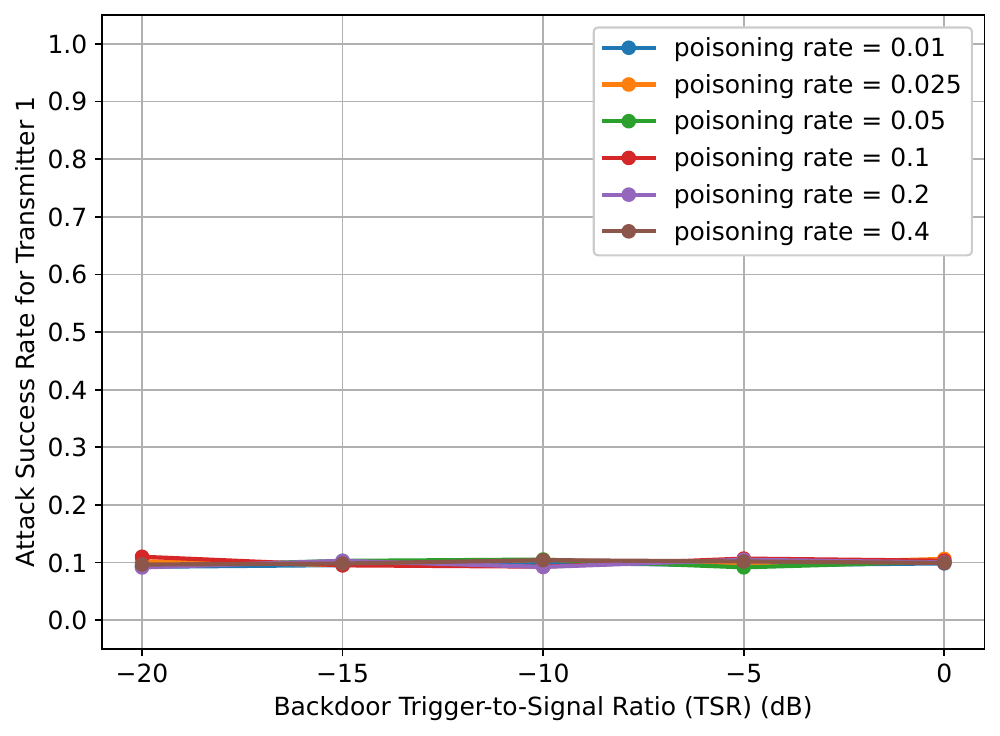}
        \caption{Transmitter~1 ASR.}
        \label{fig:defense_asr_pu}
    \end{subfigure}
    \hfill
    \begin{subfigure}{0.49\columnwidth}
        \centering
        \includegraphics[width=\linewidth, height = 0.735\linewidth]{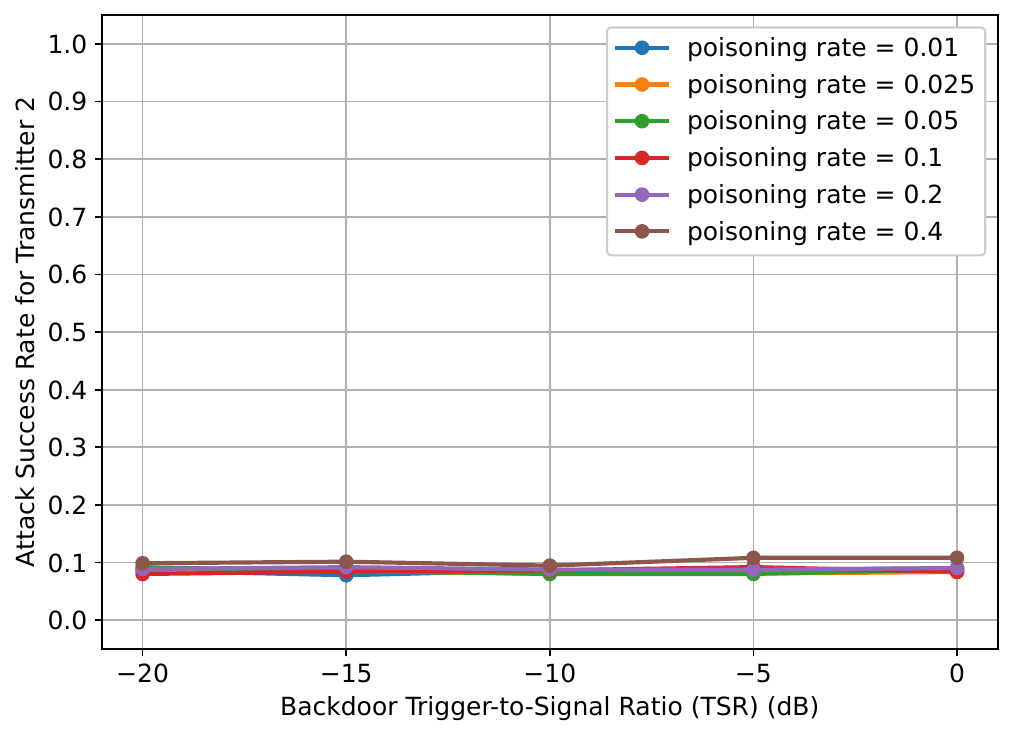}
        \caption{Transmitter~2 ASR.}
        \label{fig:defense_asr_su}
    \end{subfigure}
    \caption{Trigger-aware defense performance.} 
    \label{fig:defense_results}
    \vspace{-0.4cm}
\end{figure}

Next, we evaluate the proposed trigger-aware defense mechanism and report results in Fig.~\ref{fig:defense_results}. During defense training, trigger-contaminated observations are generated while preserving the correct semantic labels for both transmitters. The objective of the defense is to suppress the trigger-sensitive semantic features learned during backdoor poisoning while maintaining normal SemCom performance. Fig.~\ref{fig:defense_asr_su} shows that the defense significantly reduces the attack success rate for transmitter~2 across all backdoor TSRs and poisoning rates. After defense training, the attack success rate decreases to values close to random-guessing behavior, indicating that the receiver no longer strongly associates the trigger waveform with the attacker-selected target label. This shows that the trigger-aware defense successfully suppresses the trigger-conditioned semantic features introduced during backdoor poisoning. 

The semantic accuracies for both transmitters remain close to the clean no-attack performance, as shown in Figs.~\ref{fig:defense_pu_acc}-\ref{fig:defense_su_acc}. This indicates that the defense mechanism preserves the normal SemCom capability while improving robustness against adversarial trigger perturbations. The semantic inference of transmitter~1's samples remains stable after defense training, as shown in Fig.~\ref{fig:defense_asr_pu}, confirming that the defense does not introduce significant degradation to unaffected semantic users.

Averaged across all defense settings of backdoor TSRs and poisoning rates evaluated in the paper, and compared to the corresponding no-attack case, the proposed trigger-aware defense preserves the SemCom performance while effectively mitigating the backdoor attack. Averaged over transmitter~1 and transmitter~2, the semantic classification accuracy improves by \(0.32\%\). At the same time, the average reconstruction quality slightly improves by  \(1.01\)~dB in PSNR and \(9.37\%\) in SSIM relative to the unattacked baseline. These improvements may arise, in part, because the trigger-aware defense training enhances the robustness and potentially the generalization capability of the semantic decoder under perturbed wireless observations, while also regularizing the learned latent semantic representations. These results show that trigger-aware robust training is effective in protecting SemCom receivers against selective over-the-air backdoor attacks while preserving semantic reconstruction and semantic inference performance over the shared wireless channel.

\section{Conclusion}
\label{sec:conclusion}
We considered a SemCom system for simultaneous transmitters operating over a multiple access channel using semantic encoders and decoders. We introduced a selective over-the-air backdoor attack in which a low-power trigger waveform is injected into the shared wireless signal to selectively manipulate the semantic inference for one transmitter while minimally affecting the semantic performance for the other transmitter. We showed that this attack is highly effective in inducing targeted semantic misclassification for one transmitter without significantly affecting the other transmitter in shared access. To mitigate this vulnerability, we introduced a trigger-aware defense mechanism that restores semantic robustness by preserving correct semantic labels and reconstruction quality under trigger-contaminated wireless observations. The results demonstrate the vulnerability of multi-user SemCom systems to selective over-the-air backdoor attacks and highlight the importance of robust design for secure SemCom.

\bibliographystyle{IEEEtran}
\bibliography{references}

\end{document}